\documentclass[pra,aps,epsf]{revtex4}
\usepackage{graphicx}
\usepackage{epsfig}

\newcommand{\ba}{\begin{eqnarray}} 
\newcommand{\ea}{\end{eqnarray}} 
\newcommand{\be}{\begin{equation}} 
\newcommand{\ee}{\end{equation}} 
\newcommand{\bea}{\begin{eqnarray}} 
\newcommand{\eea}{\end{eqnarray}} 
\def\etal{{\it et al}.}

\def\sectref#1{Section \ref{#1}}

\def\figref#1{Figure \ref{#1}}

\begin{document}

\def\CC{{\rm\kern.24em \vrule width.04em height1.46ex depth-.07ex \kern-.30em C}}

\title{Atomistic Theory of Coherent Spin Transfer between Molecularly Bridged Quantum Dots}

\author{Joshua Schrier and K. Birgitta Whaley}

\affiliation{Department of Chemistry and Pitzer Center for Theoretical
Chemistry, University of California, Berkeley, CA 94720}

\begin{abstract}

Time-resolved Faradary rotation experiments have demonstrated coherent
transfer of electron spin between CdSe colloidal quantum dots coupled
by conjugated molecules.  We employ here a Green's function approach,
using semi-empirical tight-binding to treat the nanocrystal
Hamiltonian and Extended H\"uckel theory to treat the linking molecule
Hamiltonian, to obtain the coherent transfer probabilities from
atomistic calculations, without the introduction of any new
parameters.  Calculations on 1,4-dithiolbenzene and
1,4-dithiolcyclohexane linked nanocrystals agree qualitatively with
experiment and provide support for a previous transfer Hamiltonian
model.  We find a striking dependence on the transfer probabilities as
a function of nanocrystal surface site attachment and linking molecule
conformation.  Additionally, we predict quantum interference effects
in the coherent transfer probabilities for 2,7-dithiolnaphthalene and
2,6-dithiolnaphthalene linking molecules.  We suggest possible
experiments based on these results that would test the coherent,
through-molecule transfer mechanism.

\end{abstract}

\maketitle

\section{Introduction} \label{intro}

Colloidal semiconductor nanocrystals have attracted attention not only
because of their size tunable optical properties, but also because of the
long-lived spin coherences resulting from full three-dimensional
confinement.\cite{AK99} Potential spintronic and quantum computational
applications\cite{LD98,WAB+01} have led to the development of optical
techniques such as time-resolved Faraday rotation (TRFR) for the
measurement and manipulation of single spins in these
systems,\cite{GAP+99,Gup02,GAE+02} as well as to theoretical
methods to describe these properties.\cite{GAE+02,RER+02,SW03,CW04}
With the spin properties of isolated nanocrystals reasonably well
understood, attention has now turned to the study of spin in linked
nanocrystals.  Ouyang and Awschalom observed coherent transfer between
3.4 nm and 7.0 nm diameter colloidal CdSe coupled by
1,4-dithiolbenzene.\cite{OA03} Electron spin polarization created by
optical pumping of the larger nanocrystal is transfered
``instantaneously'' (within the experimental limitations) and coherently to the smaller nanocrystal.  This
transfer from lower band gap to larger band gap nanocrystal is
evidence against F\"orster-like transfer mechanisms.\cite{Govo03}
Ouyang and Awschalom conclude from their TRFR measurements transfer
efficiencies of $\sim 10\%$ at $T < 50 {\rm K}$ and of $\sim 20\%$ at
$75 \leq T \leq 250{\rm K}$, and attribute this temperature dependence
to changes in the conformation of the linking molecules.

Meier \etal\, have developed a transfer Hamiltonian theory
describing the TRFR signal as a function of a spin transfer
probability and Heisenberg spin-exchange between the coupled
nanocrystals.\cite{MCG+04} They draw the important conclusion that the
TRFR signal at a given probe frequency does not provide enough
information to determine the spin transfer probability, and thus the
increase in Faraday rotational signal observed experimentally---and
interpreted as an increase in spin transfer efficiency for higher
temperatures---may also result from increases in the incoherent
transfer paths.  In this paper, we describe a method for the atomistic
description of the coherent spin transfer probability, that uses a
semiempirical tight-binding theory for the treatment of the
nanocrystals and Extended H\"uckel theory for the linking molecule.
This microscopic approach allows us to calculate the coherent spin
transfer for new classes of linking molecules, as well as to study the
effects of their molecular conformation in a systematic fashion.
Additionally, we explore the variability of the spin transfer due to
the distribution of surface sites of the nanocrystal, which is a
previously unrecognized factor in the construction of devices based on
this system.

\section{Theory}

\subsection{Semiempirical Hamiltonian Calculation}\label{hamiltonian}

To treat the CdSe nanocrystals, we have used the orthogonal
nearest-neighbor $sp^{3}s^{*}$ basis tight-binding (TB) approach,
using the standard semiempirical matrix elements,\cite{LL90}
transformed from zinc blende to hexagonal crystal
structure.\cite{PW99}.  These parameters have been shown previously to
reproduce the optical\cite{LPW98} and magneto-optical\cite{SW03}
properties of this system.  The surface sites of the nanocrystal are
passivated with oxygen-like ligands to mimic the TOPO ligands used
experimentally.

Since the TB parameterization does not extend to carbon or hydrogen
atoms, we have treated the linking molecules using Extended H\"uckel
theory (EHT).\cite{Hoff63,Whangbo00} As in the TB method used for the
nanocrystal, EHT uses a basis of valence Slater orbitals.  Following
Ref. \onlinecite{PW99}, where the ionization energy of the cation was
used to determine a consistent energy scale for the CdS and CdSe TB
parameterizations in order to produce a parameterization for ${\rm
CdS_{x}Se_{1-x}}$, we require the ionization energies (diagonal TB
Hamiltonian terms) of the S and Se atoms to be consistent between the
TB and EHT schemes.  This leads to a global shift of the EHT
Hamiltonian energies by 11.155 eV to make it consistent with the TB
energy scale.  The insensitivity of our results to the exact value of
this shift is demonstrated in \sectref{14dtb_shift}.  An $sp^3$
Slater-orbital basis is used for the non-Cd atoms (H, C, S), using the
parameters of Ref. \onlinecite{Yaehmop}, and the $sp^{3}s^{*}$
Slater-orbital basis of Ref. \onlinecite{HW96} is used for the Cd
atoms.  These calculations are performed using a modified version of
the Yaehmop program.\cite{Yaehmop} For simplicity, and for better
comparison with the work of Meier \etal,\cite{MCG+04} we have
neglected the spin-orbit coupling of the linking molecule.  This may
be justified by noting that the spin-orbit coupling constants for
carbon and hydrogen atoms are small.\cite{MRW62} Nonetheless, this
could be included in future studies.\cite{PB02,PB02b} Additionally,
our treatment considers only small applied magnetic fields, such that
Zeeman splitting of the nanocrystal and molecule levels is small.

There are several possible alternatives to the present TB/EHT
Hamiltonian treatment that may be useful for future research on other
molecularly linked nanocrystal systems.  For silicon nanocrystals
linked by hydrocarbon molecules, one could use the orthogonal
nearest-neighbor tight-binding parameterizations developed for Si-C
alloys.\cite{Rob92} For metal nanocrystals, one might use the
non-orthogonal tight binding method of Papaconstantopolous and
coworkers,\cite{CMP94,MP96} combined with either EHT\cite{Hoff63} or a
Harrison tight-binding parameterization\cite{Harrison} for the linking
molecule. Additionally, one might use other semiempirical techniques
from quantum chemistry, such as INDO/S.\cite{RZ73,RZ74}

\subsection{Determination of the Nanocrystal Surface Local Density of States}
\label{ldos}

The transfer probability calculation, described
below, requires the use of the local density of states (LDOS) for the
molecular contact sites.  Unlike in typical molecular electronics
calculations, e.g., Ref. \onlinecite{KB01}, in which the final contact
is taken to be part of a bulk lead and consequently the bulk DOS at
the Fermi energy may be used, in our case the LDOS will be a function
of the nanocrystal surface site connected to the linking molecule, and
must be calculated explicitly.  By calculating the Green's function,
$G_{NC}(E)$, of the nanocrystal Hamiltonian, $H_{NC}$,

\be
 G_{NC}(E) = {\left [ \left(E+i\eta\right) I - H_{NC} \right ]}^{-1}
\ee

(where $i = \sqrt{-1}$ and $\eta$ is infinitesmal), the LDOS,
$\rho_{NC}(E,j)$, for the basis orbital $j$ is given simply by

\be
 \rho_{NC}(E,j) = - {\pi}^{-1} Im \left [ {G_{NC}(E)}_{jj} \right ],
\ee
or alternatively by
\be
g_{NC}(E,j) = Im \left [ {G_{NC}(E)}_{jj} \right ].
\ee

\noindent 

According to previous
experimental\cite{BMGB94,RNOb01,TKR01,CBK+97,DHK+02} and
theoretical\cite{PW99} work, group VI elements (such as S and O) bond
preferentially to the Cd atom surface sites on the nanocrystal.
Consequently, this calculation is made for each of the $sp^3s^{*}$ Cd
basis orbitals on the 84 surface Cd sites of the 3.4 nm diameter
nanocrystal, and for each of the $sp^3s^{*}$ Cd basis orbitals on the
243 surface Cd sites of the 5.0 nm diameter nanocrystal, over energy
intervals of $\Delta E = 0.01\,{\rm eV}$.  We use the same nanocrystal structures as in
our previous work.\cite{SW03} This stage of the calculation is time
consuming, since determining $G(E)$ requires a matrix inversion for
each value of E considered. However the results for a given
nanocrystal are independent of the linking molecule, and thus need
only be determined once and stored for future use.  We also note that
this process adds artificial peaks in the band-gap region for the
$T(E)$ plots, which are due to surface states induced by the removal
of the ligands for the calculation of $G(E)$.

\subsection{Transfer Probability Calculation}
\label{transferprob}

Following the treatment of Datta,\cite{Datta} we may express the
Green's function for the entire system as 

\be 
G_{sys}(E,j,k) = {\left
[ E I - H_{mol} - \Sigma^{R}_{NC1}(E,j) - \Sigma^{R}_{NC2}(E,k) \right
]}^{-1} ,
\ee 

\noindent where the self energy for a given nanocrystal surface site atomic
orbital (AO) $j$, is given by 

\be 
\Sigma_{NC1}(E,j) =
\tau^{\dagger}_{NC1}(j)g_{NC1}(E,j)\tau_{NC1}(j). 
\ee 

\noindent Here $\tau(j)$
is the portion of the Hamiltonian matrix between the nanocrystal surface site
AO $j$ and the linking molecule basis functions.  These matrix elements are
determined from the EHT calculation by including two Cd atoms, given
with a TB/EHT consistent $sp^{3}s^{*}$ Slater orbital basis as
described in \sectref{hamiltonian}.  Defining 

\be 
\Gamma_{NC1}(E,j) =
i\left [ \Sigma_{NC1}(E,j) - \Sigma_{NC1}^{\dagger}(E,j)\right ], 
\ee

\noindent the transmission probability is obtained as  

\be T(E,j,k) = {\rm Tr}\left
[
\Gamma_{NC1}(E,j)G_{sys}(E,j,k)\Gamma_{NC2}(E,k)G_{sys}^{\dagger}(E,j,k)
\right ]. 
\ee 

\noindent The transmission function is calculated for each of the
possible $j,k$ combinations, at energy intervals of $\Delta E = 0.01\,
{\rm eV}$.  This is then processed to determine
the statistics of $T(E)$ as a function of the surface-site
connections.  Our treatment assumes that the transfer process is entirely
coherent, and as a result may overestimate the observed transfer
probability due to neglect of incoherent loss channels.

\section{Results}
\subsection{1,4-dithiolbenzene} \label{14dtb}

\subsubsection{Surface Site Dependence} \label{14dtb_surface}

The 1,4-dithiolbenzene molecule, shown in Figure 1a, 
consists of a
benzene ring with sulfur atoms attached to opposite ends of the ring, and
is commonly used
in molecular conduction studies.\cite{KB01,KB02} 
To begin our study, we took the MM2\cite{Allinger77}
molecular-dynamics optimized structure of 1,4-dithiolbenzene, and
replaced the thiol hydrogen atoms with Cd atoms at the appropriate
bond length (2.63 \AA)\cite{Harrison}, 
as depicted in \figref{14dtb_initial_fig}. In \figref{14dtb_initial_fig}
the solid and dashed lines indicate the average and median values of
$T(E)$, respectively.  The mean, median and standard deviation of
$T(E)$ are taken over all of the $243 \times 84 = 20412$ possible
connections of the 1,4-dithiolbenzene molecule between surface sites
of the two nanocrystals. Each calculation was taken over the -1 to 6
eV energy range, at intervals of $\Delta E = 0.01\, {\rm eV}$. 
  In the region of -0.7
to 0 eV, the mean $T(E)$ oscillates between values of approximately
0.13 and 0.06; the median $T(E)$ in this region oscillates between
0.06 and 0.04.  In the region of 2.4 to 3 eV, both the mean and the
median $T(E)$ reach a peak at 2.6 eV, of value 0.30 and 0.21,
respectively.  Prior to this peak, the mean $T(E)$ ranges from 0.04 to
0.18 in the region of 2.4 to 2.6 eV, and from 0.10 to 0.08 in the
region from 2.6 to 3 eV; the median $T(E)$ follows the same
qualitative behavior, but with lower values.  The mean transmission
probabilities in both of these energy ranges, corresponding to the
hole and electron transmission probabilities, are similar to those
determined by Meier \etal\, from their analysis of the TRFR
data.\cite{MCG+04} Note that the latter was non-microscopic and
employed a fit to the experimental results.

\subsubsection{Role of the Energy Scale Shift} \label{14dtb_shift}

As mentioned in \sectref{hamiltonian}, the absolute energy scales of
the TB and EHT Hamiltonians are not the same, requiring an energy
shift to make them consistent.  By calculating the average difference
between the S and Se atomic orbital ionization potentials (diagonal
Hamiltonian matrix elements) obtained in the two different methods,
one obtains an average shift of 11.155 eV.  To explore the dependence of our
results on the exact value of this energy correction, we performed
calculations for T(E) using shift corrections in the range of 10.0 to
12.0 eV, at intervals of 0.5 eV, using the same molecular geometry as
in the previous section.  We found that the changes in mean $T(E)$ due
to the different choices of energy shift lie within the variation due to surface
site attachment for calculations madee with a shift of 11.155 eV, with
the exception of some peaks in the nanocrystal band gap region and in
the region below -0.75 eV (in the TB parameterization, 0 eV
corresponds to the highest occupied molecular orbital).  These results
indicate that the exact value of this shift is unimportant for band
edge electrons and holes, compared to the effects of variation due to
the surface site attachment of the molecule.

\subsubsection{Conformational Dependendence} \label{14dtb_conformation}

Previous theoretical studies of thiolated molecules between gold
contacts by Kornilovitch and Bratkovsky indicate a strong dependence
of the transmission function on the conformation of the molecule, due
to the directionality of the $p$-type atomic orbital interactions
between molecule and contact atoms.\cite{KB01,KB02} This
conformational dependence was conjectured by Ouyang and Awschalom to
be responsible for the temperature dependence of the spin transfer
observed in the experimental study.\cite{OA03} This is straightforward
to study within the current microscopic theory.  We first examined the
case in which both of C-S and S-Cd bonds are collinear in the plane of
the benzene ring, shown in \figref{14dtb_flat_fig}.  Such conformation
shows a decrease in both the hole and electron region $T(E)$ values
below $\sim 0.05$.  Next we examined the effects of moving the S-Cd
bond in the plane and out of the plane of the benzene ring, shown
respectively in Figures 4 and 5.  For the in-plane motions, the value of
$T(E)$ below 2.5 eV remains relatively constant, and the region
between 4-5 eV shows the greatest changes.  When one of the Cd-S bonds
was approximately collinear with the S-C bond, the effect due to
moving the other bond was negligible compared to the case of both Cd-S
colinear with their respective S-C bonds.  Bends of both C-S-Cd
linkages were required to obtain an enhancement in $T(E)$.  For the
out of benzene plane motions, shown in \figref{14dtb_outplane_fig}, we
found enhancements in $T(E)$ in the range of -1 to +0.5 eV, with
changes in the 4 to 5 eV range only for the extreme angles.  As seen in
the figure, the $T(E)$ for the hole reached a maximum of $\sim 0.25$.
We note that the transmission probability for the hole is dependent on
the out of plane component of the C-S-Cd dihedral angle, while the
electron transmission probability is dependent on the in-plane angle.

\subsection{1,4-dithiolcyclohexane} \label{14dtch}

As shown in Figure 1b, 1,4-dithiolcyclohexane differs from
dithiolbenzene by the absence of pi-bonding between carbon atoms in
the ring (due to bonding with additional hydrogen atoms making it a
saturated molecule), which in turn decreases its conductivity.  Just
as for the 1,4-dithiolbenzene discussed in \sectref{14dtb_surface}, we
began with the MM2 optimized structure for 1,4-dithiolcyclohexane.
The results for this initial conformation (the so-called ``boat''
conformation\cite{MorrisonBoyd} of cyclohexane) are shown in
\figref{14dtch_initial_fig}.  In the region of -1 to 0 eV and 2 to 3
eV we found the mean $T(E)$ to be less than $0.01$, with the exception
of peaks at 2.6 eV and 2.8 eV, where $T(E)$ rises to $\sim 0.04$.  We
then performed all of the C-S-Cd rotations that were carried out in
\sectref{14dtb_conformation} for 1,4-dithiolbenzene. The mean $T(E)$
determined in each of these calculations is plotted as a separate line
in \figref{14dtch_rotation_fig}.  From this it is evident that $T(E)$
for this saturated linking molecule is insensitive to conformational
changes in the valence and conduction band edge energy regions,
remaining near zero for all cases.  This is consistent with the
experimental results reported by Ouyang and Awschalom.\cite{OA03} The
comparison to 1,4-dithiolcyclobenzene is discussed in detail in
\sectref{disc_comparison}.

\subsection{2,6-dithiolnaphthalene} \label{cisnap}

Dithiolnaphthalene consists of two aromatic rings joined at an edge.
We considered two placements of the thiol groups on this ring
structure in order to investigate the effect of length of the aromatic
linker and the posibility of quantum interference effects.  The first
placement corresponds to 2,6-dithiolnaphthalene, shown in Figure 1c.
The ring structure was MM2 optimized, and the C-S-Cd conformation was
set to the maximal $T(E)$ conformation determined in
\sectref{14dtb_conformation} for 1,4-dithiolbenzene.  Due to the
higher computational costs as a result of the larger molecule size, we
examined only the cases depicted in the cartoon in
\figref{cisnap_fig}, in which both C-S-Cd dihedral angles with respect
to the plane of the naphthalene ring were oriented in the same
direction and in the opposite direction (denoted ``up, up'' and ``up,
down'', respectively), and the case of the C-S-Cd angle in the plane
of the ring (denoted ``flat'').  As shown in \figref{cisnap_fig},
$T(E)$ in the regions of -1 to 0 eV and 2.0 to 3.5 eV are near zero
for all of the conformations we examined.  In the energy range between
4 to 5 eV, the bent conformation with both C-S-Cd units oriented in
the same direction was found to have a peak $T(E)$ of 0.07, as
compared to 0.12 for the conformation with the C-S-Cd units in
opposite directions.

\subsection{2,7-dithiolnaphthalene} \label{transnap}

We examined the analogous conformations as described in the previous
section for the 2,7-dithiolnaphthalene linked nanocrystals, shown in
Figure 1d, and in the cartoons in \figref{transnap}.  
As shown in \figref{transnap_fig}, $T(E)$ in the region of
-1 to 0 eV varies between $0.09$ and $0.02$ for the bent
conformations, and $0.01$ to zero for the flat conformation.  In the
2.0 to 4.0 eV range, all mean $T(E)$ values are near zero.  Both
conformations have similar mean $T(E)$ values over the entire energy
range, and unlike the 2,6-dithiolnaphthalene discussed above, the
variation in $T(E)$ due to surface site attachment was also found to
be similar for both bent conformations.  A detailed analysis of the
difference in $T(E)$ for the two substituted naphthalene molecules is
presented in \sectref{disc_interference} below.

\section{Discussion}

\subsection{Surface Site Dependence}

The variations in $T(E)$ due to the specific surface site attachments,
discussed in \sectref{14dtb_surface}, and observed in all of the
calculations described subsequently, presents a quality-control
problem for the construction of individual spin-manipulating devices
such as required for quantum computation.  While the ensemble of
molecularly linked nanocrystals observed in optical experiments will have a
non-negligible spin transfer percentage, any particular device may
have much better (or much worse) performance.  However, there may be
chemical effects during the assembly process which preferentially
select a subset of surface sites for binding to the molecule.
Alternately, use of highly symmetric nanocrystals, such as the icosahedral Si
nanocrystals recently predicted by Zhao \etal,\cite{ZKDZ04} may also simplify
the problem by reducing the number of possible surface sites available
for attachment.  In any case, we conclude that it is essential to understand the effect
of variability in surface site linkage on the coherent spin transfer probability.

\subsection{Comparison of Theoretical and Experimental Results for
1,4-dithiolbenzene and 1,4-dithiolcyclohexane}\label{disc_comparison}

The results presented in \sectref{14dtb} with \sectref{14dtch} agree
with the measurements of Ouyang and Awschalom (10-20\% transfer for
dithiolbenzene, no transfer for dithiolcyclohexane) \cite{OA03} and
corresponding estimates by Meier \etal\, (6-13\% for
dithiolbenzene)\cite{MCG+04}.  Unlike the latter study, we have
neglected the role of the coulomb contribution due to charging of the
QDs, and so our computed forward and backware transfer probabilities
are equivalent.  Furthermore, since the estimate of the transfer probability
in Meier \etal\, is derived from a fitting to experimental
data, it implicitly performs an average over conformations. As we discuss
in the next section, this can lead to changes in the transfer
probability. Consequently we can only expect qualitative agreement
from calculations for a single conformation.  Nevertheless, our
calculation of negligible valence and conduction carrier transfer for
the dithiolcyclohexane linked region shows the qualitative benefit of
the atomistic approach, and suggests that this approach may be useful
to provide parameters for model studies in future work.

\subsection{Conformation as a Mechanism for Temperature Dependence}

Our results results for the conformational dependence of $T(E)$, shown
in Figures 3-5, are consistent with the explanation of temperature
dependence of TRFR put forward by Ouyang and Awschalom,\cite{OA03}
namely, that different molecular conformations populated as a function
of temperature can exhibit different spin transfer properties.  In
addition to conformational dependence, 
thermally populated vibrational modes may play a role in the temperature
dependence.  We have neglected explicit coupling of the electronic
modes to vibrational modes here, although in principle the method may
be extended to treat this.\cite{Datta,CZdV04,CTD04} We suggest that
this may be studied experimentally by means of
isotopically-substituted linking molecules.  If the temperature
dependence is primarily due to the molecular conformation, the
isotopic substitutions will have only a small effect on the observed
spin transfer.  Alternately, if the temperature dependence is due to
population of some subset of vibrational states, selectively
isotopically substituted linker molecules could then be used to
identify which vibrational modes play a dominant role, by making use of
isotope effects on vibrational frequencies.

\subsection{Quantum Interference Effects}\label{disc_interference}

Baer and Neuhauser demonstrated that the phase coherence (or
anticoherence) through a molecular ring can be used as a molecular XOR
switch.\cite{BN02,BN02a} In a simple H\"uckel model at half-filling,
the wavefunction changes sign at every second site, giving rise to
coherent transfer when the lengths of the two complementary 
portions of the ring differ by $2n$ sites, and resulting in ``anti-coherent'' transfer when
the corresponding lengths differ by $2n+2$ sites.  This effect has also been
demonstrated in density-functional calculations of polycyclic
hydrocarbon molecules.\cite{WNB04} In \sectref{cisnap} and
\sectref{transnap} we predict that this difference between coherent
and anticoherent transfer for different loop lengths should be visible
in the substituted naphthalene-linked nanocrystals.  In terms of the
H\"uckel model described above, this can be understood by counting the
number of carbon atoms between the thiol groups 
for the naphthalene molecules shown in Figure
1c and 1d,  traversing the upper and lower portions of the ring.
  For
the 2,6-dithiolnaphthalene molecule shown in Figure 1c, there is a
difference of two sites for the upper and lower paths,
resulting in destructive interference.  For the 2,7-dithiolnaphthalene
molecule shown in Figure 1d, the lengths of the 
upper and lower paths between the
thiol groups is the same, resulting in constructive interference.
This suggests an experimental test of the coherent transport mechanism
proposed by Meier \etal,\cite{MCG+04} that was taken as a given ansatz
in this work.  Specifically, an observation of significant 
spin-transfer between 
2,7-substituted napthalene linked nanocrystals without significant transfer for
 2,6-substituted naphthalene linked nanocrystals would provide
positive evidence that hole conduction occurs through the molecule.
Furthermore, we note that extending the distance between the
nanocrystals to approximately 12.9 \AA\, as is the case with the
naphthalene linker, and finding persistence of substantial 
spin transfer, would provide further evidence against the
possibility of F\"orster-like\cite{Govo03} spin transfer mechanisms.

\section{Summary}

We have developed an atomistic semi-empirical theory for the coherent
transfer of carriers between molecularly linked nanocrystals.  Our
results for 1,4-dithiolbenzene and 1,4-dithiolcyclohexane linked
nanocrystals are in qualitative agreement with both
experimental\cite{OA03} and empirical model theoretical\cite{MCG+04}
work on these systems.  Using an atomistic approach, based on a
semiempirical tight-binding description of the nanocrystals, and
extended H\"uckel theory description of the linker molecules, and a
Green's function formalism for the description of the transfer
probability, we examined the role of surface site attachment and
linking molecule conformation on the coherent transfer probability.
The variation due to the former is a new insight which may be
overlooked in ensemble measurements; our calculations indicate a
standard deviation in $T(E)$ due to surface site attachments of
0.1-0.2.  The variation due to the latter supports the hypothesis of
Ouyang and Awschalom that the temperature dependence of the spin
transfer results from conformational changes in the linking molecule.
In addition, we have predicted a quantum interference effect in the
coherent transfer probabilities for 2,6-dithiolnaphthalene and
2,7-dithiolnaphthalene linked nanocrystals, which we propose be used
as an experimental test of the coherent nature of the transfer.

\section{Acknowledgements}

J.S. thanks the National Defense Science and Engineering Grant (NDSEG)
program and U.S. Army Research Office Contract/Grant
No. FDDAAD19-01-1-0612 for financial support.  
This work was also supported by the
Defense Advanced Research Projects Agency (DARPA) and the Office of
Naval Research under Grant No. FDN00014-01-1-0826, and the National
Science Foundation under Grant EIA-020-1-0826.


\vfill\eject

\begin{figure}
\includegraphics[width=8.5cm]{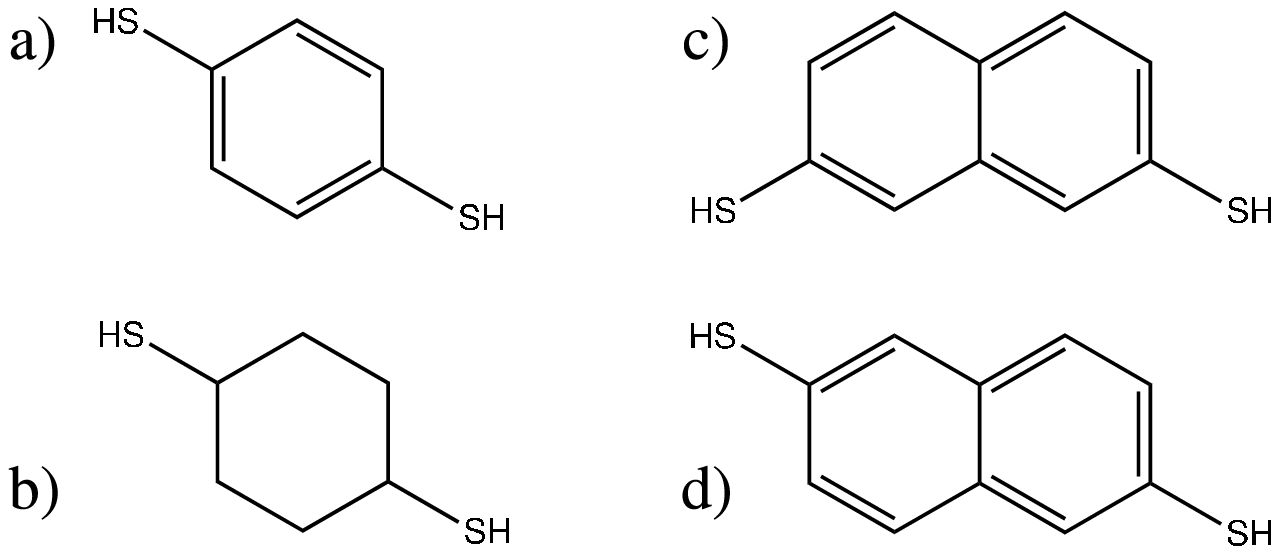}
\caption{Linking molecules treated in this study: a) 1,4-dithiolbenzene; b) 1,4-dithiolcyclohexane; c) 2,6-dithiolnaphthalene; d) 2,7-dithiolnaphthalene.}
\label{molecules_fig}
\end{figure}

\begin{figure}
\includegraphics[width = 8.5cm]{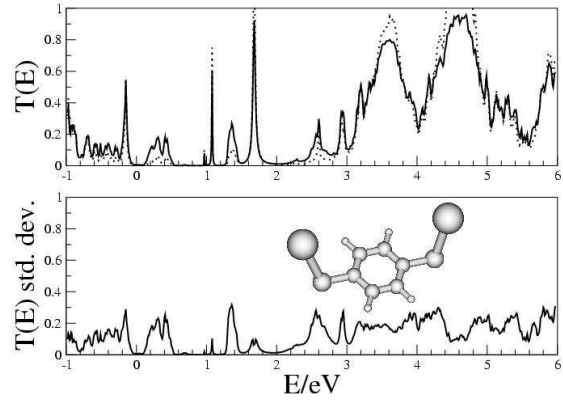}
\caption{$T(E)$ (top) and standard deviation of $T(E)$ (bottom) for
1,4-dithiolbenzene linked CdSe nanocrystals of 3.4 nm and 5.0 nm
diameter.  The MM2 optimized conformation of the molecule used in this
calculation is shown as the inset cartoon.  The black solid and black
dashed lines indicate the average and median $T(E)$, respectively.
The standard deviation of $T(E)$ results from variations due to the
molecule-nanocrystal surface site attachments, as described in
\sectref{14dtb_surface}. The mean value of $T(E)$ in the conduction
band region (-0.7 - 0 eV) oscillates between 0.06-0.13.  Peaks in
$T(E)$ in the bandgap region (0-2.4 eV) results from artificial
surface states due to ligand removal in the calculation of $G(E)$.}
\label{14dtb_initial_fig}
\end{figure}


\begin{figure}
\includegraphics[width = 8.5cm]{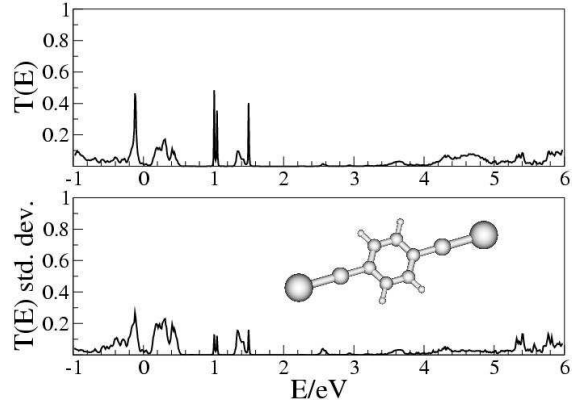}
\caption{$T(E)$ for the flat conformation of the 1,4-dithiolbenzene
linker molecule, in which both the C-S and S-Cd bond are collinear in
the plane of the benzene ring.  
The upper figure shows mean $T(E)$ and the lower figure the standard
deviation, due to the variability in surface site attachment.
As compared to the reference conformation in
\figref{14dtb_initial_fig}, the hole energy region transfer
probabilities are reduced by half, and the electron energy region
$T(E)$ is nearly zero.}
\label{14dtb_flat_fig}
\end{figure}

\begin{figure}
\includegraphics[width = 8.5cm]{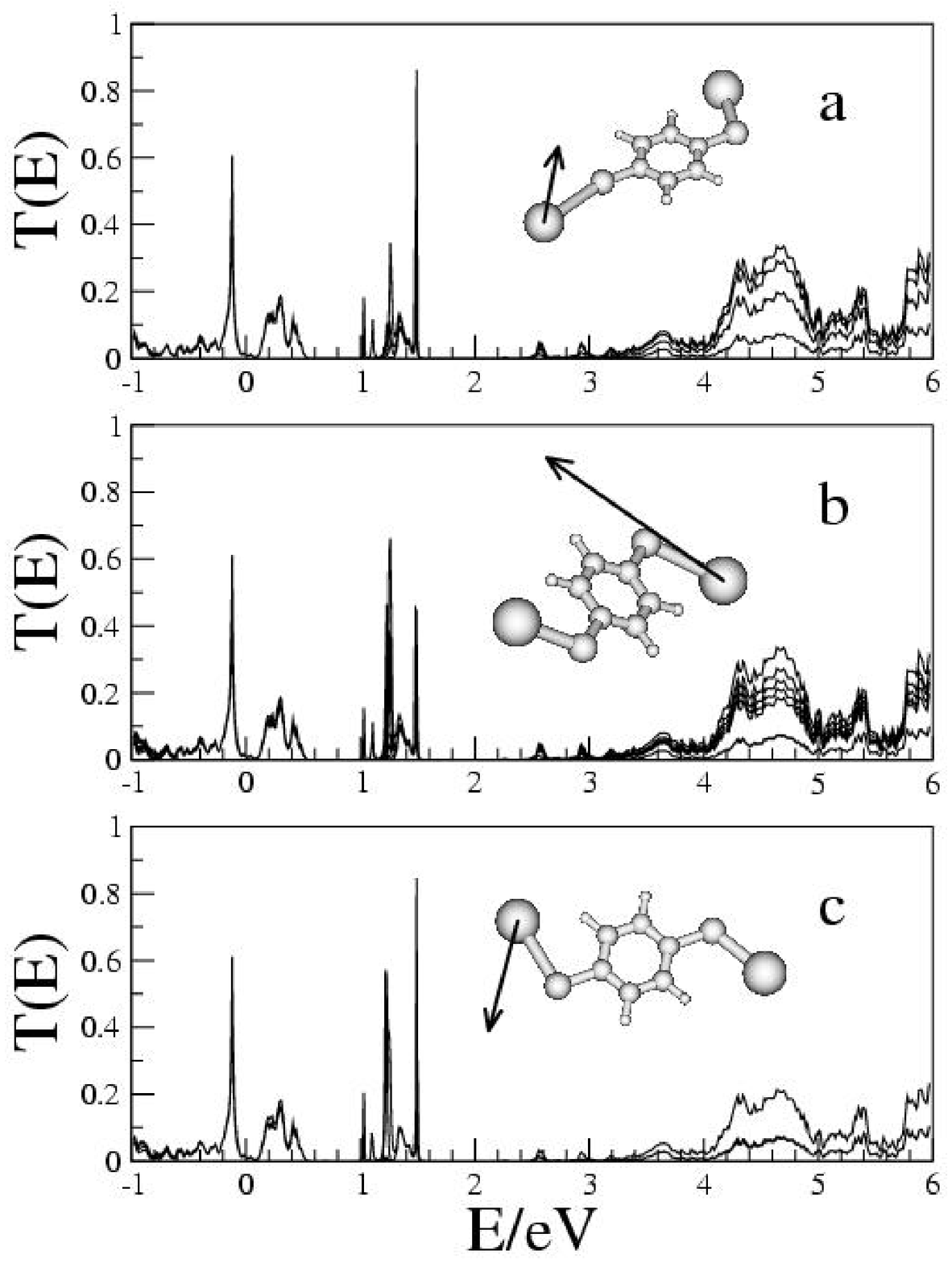}
\caption{Effect on $T(E)$ of in-plane bond angle changes from the flat
conformation of 1,4-dithiolbenzene linked nanocrystals.  Arrows from
the molecule indicate the direction of the motion studied; each line
in the plots indicates a different conformation along that
direction.
}
\label{14dtb_inplane_fig}
\end{figure}

\begin{figure}
\includegraphics[width = 8.5cm]{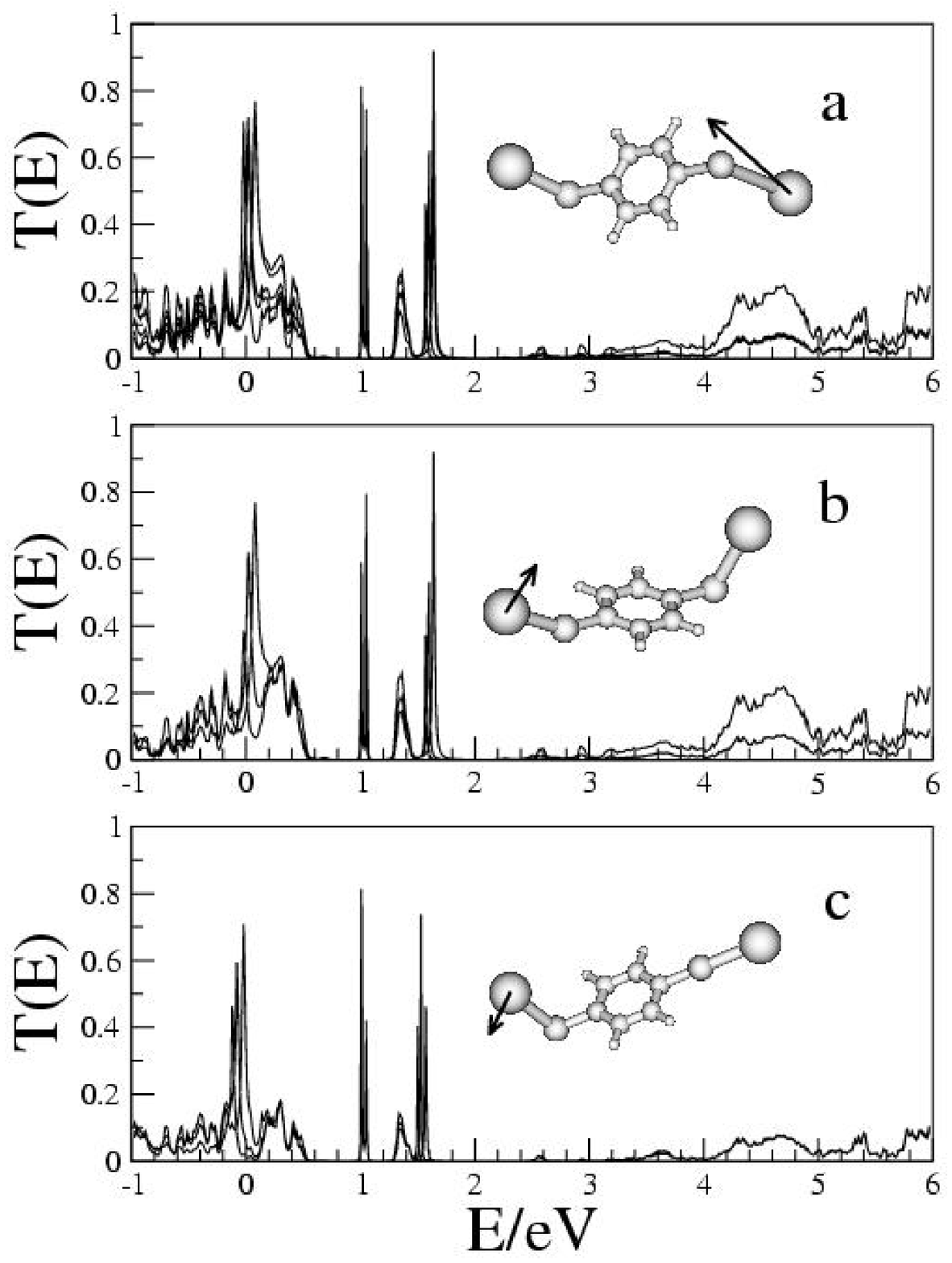}
\caption{ Effect on $T(E)$ of out-of-plane bond angle changes from the flat conformation of 1,4-dithiolbenzene linked nanocrystals.  Arrows from
the molecule indicate the direction of the motion studied; each line
in the three plots indicates a different conformation along that
direction.
}
\label{14dtb_outplane_fig}
\end{figure}

\begin{figure}
\includegraphics[width = 8.5cm]{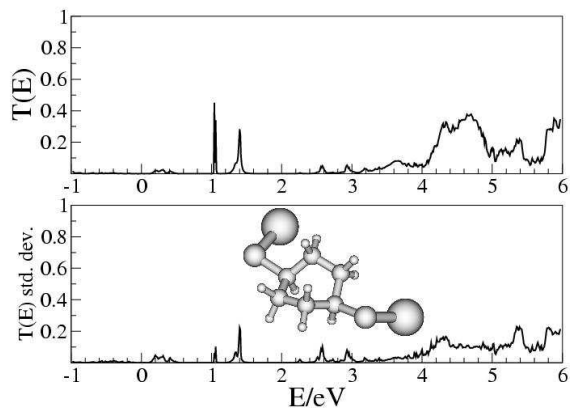}
\caption{Mean $T(E)$ and standard deviation for the MM2-optimized
(``boat'') conformation of
1,4-dithiolcyclohexane linked nanocrystals.  
Note the dramatic
decrease in $T(E)$ over the entire energy range as compared with the
conjugated linking molecule 1,4-dithiolbenzene, shown in
\figref{14dtb_initial_fig}.}
\label{14dtch_initial_fig}
\end{figure}

\begin{figure}
\includegraphics[width = 8.5cm]{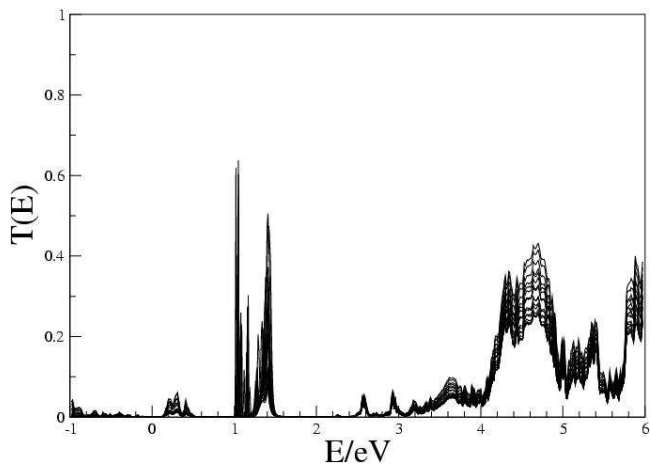}
\caption{Effect of C-S and S-Cd bond rotations on the transmission
probability, $T(E)$, for 1,4-dithiolcyclohexane.
The results for each C-S and S-Cd rotation calculated as described in
\sectref{14dtch} are shown as individual lines in the figure.
No significant changes in $T(E)$ are observed for any of these
rotational motions.}
\label{14dtch_rotation_fig}
\end{figure}

\begin{figure}
\includegraphics[width = 8.5cm]{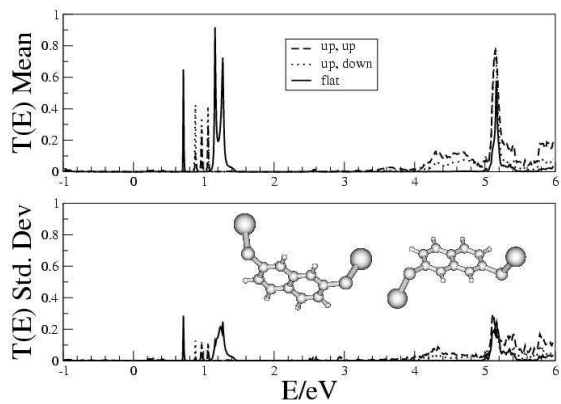}
\caption{Mean $T(E)$ and surface site standard deviation for 
2,6-dithiolnaphthalene linked nanocrystals.  In the ranges of $E=-1\to 0
{\rm eV}$ and $E = 2.0\to 3.5 {\rm eV}$, $T(E)$ was found to be near zero for all
three conformations.  This results from destructive quantum interference of
the electron paths, as discussed in detail in \sectref{disc_interference}.   }
\label{cisnap_fig}
\end{figure}

\begin{figure}
\includegraphics[width = 8.5cm]{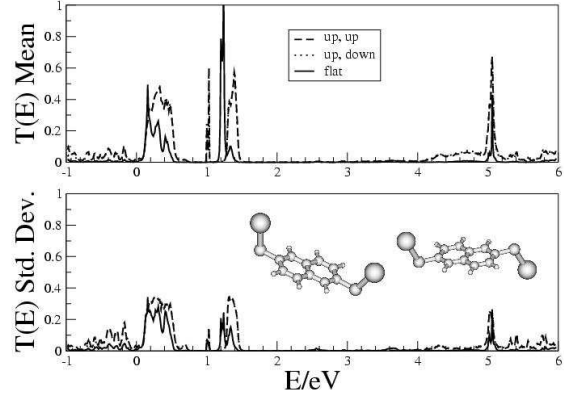}
\caption{Mean $T(E)$ and surface site standard deviation for 
2,7-dithiolnaphthalene linked nanocrystals. 
In contrast to the result for 2,6-dithiolnaphthalene shown in
\figref{cisnap_fig}, constructive quantum interference of the electron
paths causes a non-zero $T(E)$ of between $0.02$ to $0.09$ in the
region of $E = -1 \to 0 {\rm eV}$ that is evident for all three conformations.}
\label{transnap_fig}
\end{figure}

\end{document}